\documentclass[11pt]{article}
\usepackage{graphicx}
\usepackage{amssymb}
\usepackage{amscd}
\usepackage{mathrsfs}
\usepackage{longtable,lscape}
\usepackage{amsthm}
\usepackage{amsfonts}
\usepackage{amsmath}
\usepackage{bbm}
\usepackage{float}
\usepackage{url}

\newcommand{\real}{{\mathbb R}}
\newcommand{\captionfonts}{\footnotesize}
\makeatletter 
\long\def\@makecaption#1#2{%
\vskip\abovecaptionskip
\sbox\@tempboxa{{\captionfonts #1: #2}}%
\ifdim \wd\@tempboxa >\hsize
{\captionfonts #1: #2\par}
\else
\hbox to\hsize{\hfil\box\@tempboxa\hfil}%
\fi
\vskip\belowcaptionskip}
\makeatother 
\begin{document}
\title{A possible solution to the second entanglement paradox}
\author{Diederik Aerts$^1$ and Massimiliano Sassoli de Bianchi$^{2}$ \vspace{0.5 cm} \\ 
\normalsize\itshape
$^1$ Center Leo Apostel for Interdisciplinary Studies \\
\normalsize\itshape
Brussels Free University, 1050 Brussels, Belgium \\ 
\normalsize
E-Mail: \url{diraerts@vub.ac.be}
\\ 
\normalsize\itshape
$^2$ Laboratorio di Autoricerca di Base \\ 
\normalsize\itshape
6914 Lugano, Switzerland \\
\normalsize
E-Mail: \url{autoricerca@gmail.com} \\
}
\date{}
\maketitle
\begin{abstract}
\noindent 
Entangled states are in conflict with a general physical principle which expresses that a composite entity exists if and only if its components also exist, and the hypothesis that pure states represent the actuality of a physical entity, i.e., its `existence'. A possible way to solve this paradox consists in completing the standard formulation of quantum mechanics, by adding more pure states. We show that this can be done, in a consistent way, by using the extended Bloch representation of quantum mechanics, recently introduced to provide a possible solution to the measurement problem. Hence, with the solution proposed by the extended Bloch representation of quantum mechanics, the situation of entangled states regains full intelligibility.
\\

\noindent PACS numbers: 03.65.Ud, 03.65.Ta
\end{abstract}
\medskip
{\bf Keywords}: Entanglement; Density operators; SU(N); Extended Bloch representation.
\\

Since their discovery, entangled states were considered to describe enigmatic situations.  More precisely, we can distinguish two entanglement paradoxes. The first one is well-known, as it attracted most of the attention of physicists throughout the years: it is related to the experimental fact that two entangled entities are able to produce perfect correlations, even when separated by large spatial distances. In other terms, the first entanglement paradox is about the existence of \emph{non-local} effects, allowing entities to become spatially separated, by arbitrarily large distances, without becoming experimentally separated. 

The second entanglement paradox is
in our opinion even more puzzling, although lesser known. Schr{\oe}dinger, the discoverer of entanglement, was perfectly aware of this difficulty, for instance when he emphasized that for two quantum entities in an entangled state only the properties of the pair appeared to be defined, whereas the individual properties of each one of the two entities forming the pair remained totally undefined~\cite{Schroedinger1935} (see also~\cite{Fraassen1991}, Sec.~7.3, and the references therein). 

Let us explain why this observation is able to generate a paradox. For this, we start by enouncing two very general physical principles (GPP)~\cite{Aerts2000}:

\begin{quotation}
\vspace{0.50cm}
\noindent {\bf GPP1} \emph{A physical entity $S$ is said to exists at a given moment, if and only if it is in a pure state at that moment.}

\vspace{0.10cm}
\noindent {\bf GPP2} \emph{A composite physical entity $S$, formed by two sub-entities $S^A$ and $S^B$, is said to exist at a given moment if and only if $S^A$ and $S^B$ exist at that moment.}
\end{quotation}

\vspace{0.50cm}
\noindent On the other hand, according to standard quantum mechanics (SQM), the following two principles are also assumed to hold:

\begin{quotation}
\vspace{0.50cm}
\noindent {\bf SQMP1} \emph{If $S$ is a physical entity with Hilbert space ${\cal H}$, each ray-state of ${\cal H}$ is a pure state of $S$, and all the pure states of $S$ are of this kind.}

\vspace{0.10cm}
\noindent {\bf SQMP2} \emph{The pure states of a composite entity $S$, formed by two sub-entities $S^A$ and $S^B$, with Hilbert spaces ${\cal H}^A$ and ${\cal H}^B$, are the rays of ${\cal H} ={\cal H}^A\otimes {\cal H}^B$. $S^A$ and $S^B$ are in the ray-states $|\psi^A\rangle$ and $|\phi^B\rangle$ if and only if $S$ is in the product ray-state $|\psi^A\rangle\otimes|\phi^B\rangle$.}
\end{quotation}

\vspace{0.50cm}
\noindent
The paradox results from the observation that the above four principles are not compatible with each other, precisely because of the existence of entangled states. Indeed, if the composite entity $S$ is in the entangled state:
\begin{equation}
|\psi\rangle = a_1\, e^{i\alpha_1}|\psi^A\rangle\otimes |\phi^B\rangle + a_2\, e^{i\alpha_2}|\phi^A\rangle\otimes |\psi^B\rangle,
\label{entanglement}
\end{equation}
with $a_1,a_2,\alpha_1,\alpha_2\in\real$, $0\leq a_1,a_2 \leq 1$, $a_1^2+a_2^2 =1$, $|\psi^A\rangle,|\phi^A\rangle \in {\cal H}^A$, $|\psi^B\rangle,|\phi^B\rangle \in {\cal H}^B$, $\langle\psi^A|\phi^A\rangle = \langle\psi^B|\phi^B\rangle = 0$, considering that $|\psi\rangle$ is a ray-state, by the SQMP1 it describes a pure state of $S$. By the GPP1, we know that $S$ exists, and by the GPP2, the two sub-entities $S^A$ and $S^B$ also exist. But then, by the SQMP1, $S^A$ and $S^B$ are in ray-states, and by the SQMP2, $S$ is in a product state, which is a contradiction.

Facing this conflict between the above four principles, a possible strategy is that of considering that the GPP2 does not have general validity, in the sense that when a composite entity is in an entangled state, its sub-entities would simply cease to exist, in the same way that two water droplets cease to exist when fused into a single larger droplet. However, this strategy is not fully consistent, as two quantum entities do not completely disappear when entangled, considering that there are properties associated with the pair that remain always actual. 

For instance, we are still in the presence of two masses, which can be separated by a large spatial distance. So, entanglement is neither a situation where two masses are completely fused together, nor a situation where a spatial connection would bond them together, making it difficult to spatially separate them (as in chemical bonds). Also, entangled entities can remain perfectly correlated, and the property of ``being perfectly correlated'' is clearly a property which is meaningful only when we are in the presence of two \emph{existing} entities.

In other terms, we cannot affirm that in an entangled state the composing entities cease to exist,
and it is very difficult to even conceive a situation where the GPP2 would cease to apply. 
Even the above example of two droplets of water fused together cannot be considered as a counterexample, as in this case we are not really allowed to describe the larger droplet, once formed, to be the combination of two \emph{actual} sub-droplets. So, it doesn't seem reasonable to abandon the GPP2, and 
of course it is neither reasonable to abandon the GPP1, which is almost tautological if we consider that a pure state, by definition, describes the objective condition in which an entity \emph{is} in a given moment, precisely because of its \emph{actual} existence. 

Regarding the GPP1, it could be objected that also an entity which is not in a pure state can exist. This is correct, but we have here to understand that the term `pure state' is a sort of pleonasm, as if we consider the notion of state at its most fundamental level, i.e., as a description of the \emph{reality of an entity}, in a given moment, then all states are by definition pure states. Indeed, the non-pure states (i.e., the statistical mixtures of states), only describe our subjective lack of knowledge regarding the actual condition of the physical entity under consideration. Even when we are ignorant about the objective condition of an entity, i.e., its pure state, we can of course still say that the entity exists. However, and this is how the GPP1 should be understood, if the entity exists, in a given moment, then it must be possible (at least in principle, and independently of our subjective `state of knowledge') to characterize its potential behavior under all possible interactions, and such characterization is precisely what a pure state is all about.    

So, the only way to resolve the paradox seems to be that of revisiting the SQM1. To do so, we start by observing that there is a well-defined procedure in SQM that allows to associate individual states to entangled sub-entities. If, say, we are only interested in the description of $S^A$, irrespective of its correlations with $S^B$, all we have to do is to take a \emph{partial trace}. For this, one has to rewrite the ray-state (\ref{entanglement}) in operatorial form. Defining: $D_\psi=|\psi\rangle\langle \psi|$, $D_{\psi}^A=|\psi^A\rangle\langle \psi^A|$, $D_{\phi}^A=|\phi^A\rangle\langle \phi^A|$, $D_{\psi}^B=|\psi^B\rangle\langle \psi^B|$ and $D_{\phi}^B=|\phi^B\rangle\langle \phi^B|$, we have:
\begin{equation}
D_\psi =a_1^2\, D_{\psi}^A \otimes D_{\phi}^B + a_2^2\, D_{\phi}^A\otimes D_{\psi}^B + I^{\rm int},
\label{entanglement-operator}
\end{equation}
where the interference contribution is given by:
\begin{equation}
I^{\rm int} = a_1a_2\, e^{-i\alpha}|\psi^A\rangle\langle \phi^{A}|\otimes |\phi^{B}\rangle\langle \psi^{B}| + {\rm c.c.},
\label{interference-contribution}
\end{equation}
with $\alpha =\alpha_2-\alpha_1$. The state of $S^A$, irrespective of its correlations with $S^B$, can then be naturally defined by taking the partial trace: $D^A = {\rm Tr}_B\, D_\psi$, and similarly for $S^B$: $D^B = {\rm Tr}_A\, D_\psi$. A simple calculation yields:
\begin{equation}
D^A =a_1^2\, D_{\psi}^A + a_2^2\, D_{\phi}^A, \quad D^B =a_1^2\, D_{\phi}^B + a_2^2\, D_{\psi}^B.
\label{individual states}
\end{equation}

However, (\ref{individual states}) cannot be considered to solve the second entanglement paradox, as is clear that the \emph{reduced} one-entity states $D^A$ and $D^B$ will not in general be ray-states, but \emph{density operators}, and by the SQM1 we cannot interpret them as pure states. Therefore, we cannot use the GPP1 to decree the existence of $S^A$ and $S^B$. 

The following questions then arise: To save the intelligibility of the entangled states, shouldn't we complete the SQM by also allowing density operators to describe pure states? And more importantly: Do we have sufficient physical arguments to consider such a \emph{completed quantum mechanics} (CQM), and can it be formulated in a sufficiently general and consistent way? It is the purpose of this article to provide positive answers the above questions, showing that the second entanglement paradox can be solved.

For this, we start by observing that the first reason to consider that density operators should also describe pure states is precisely the existence of the above mentioned partial trace procedure. Indeed, there is no logical reason why by focusing on a component of a system in a well-defined pure state, taking a partial trace, we would suddenly become ignorant about the condition of such component. 

Another important reason is that a same density operator admits infinitely many representations as a mixture of one-dimensional projection operators~\cite{Hughston1993}. This immediately suggests that the mixture interpretation is generally inappropriate, not only because it remains ambiguous, but also, and especially, because it fails to capture the dimension of potentiality that a density operator is able to describe. 

Another relevant observation is that composite entities in ray-states can undergo unitary evolutions such that the evolution inherited by their sub-entities will make them continuously go from ray-states to density operator states, and return; a situation hardly compatible with the statistical ignorance interpretation of the density operators (see \cite{Beltrametti1981}, sect. 7.5).

But we think there is an even more important reason to consider that the density operators can also describe pure states: if we do so, it becomes possible to derive the \emph{Born rule} and provide a solution to the measurement problem, as recently demonstrated in what we have called the \emph{extended Bloch representation of quantum mechanics}~\cite{AertsSassoli2014c}. 

Let us briefly explain how this works. As is known, the ray-states of two-dimensional systems (like spin-${1\over 2}$ entities) can be represented as points at the surface of a 3-dimensional unit sphere, called the \emph{Bloch sphere}~\cite{Bloch1946}, with the density operators being located inside of it. What is less known is that a similar representation can be worked out for general $N$-dimensional 
systems~\cite{Kimura2003}.
The 3-d Bloch sphere is then replaced by a $(N^2-1)$-dimensional unit sphere, with the difference that, for $N>2$, only a convex portion of it is filled with states.

When this generalized Bloch sphere representation is adopted, as an alternative way to represent the quantum states, it can be further extended by also including the measurements. These are geometrically described as $(N-1)$-simplexes inscribed in the sphere, whose vertices are the eigenvectors of the measured observables. These measurement simplexes, in turn, can be viewed as abstract structures made of an unstable and elastic substance, and it can be shown that an ideal quantum measurement is a process where the abstract point particle representative of the state first plunges into the sphere, in a deterministic way, along a path orthogonal to the simplex, then attaches to it, and following its indeterministic disintegration, and consequent collapse, is brought to one of its vertices, thus producing the outcome of the measurement, in a way that is perfectly consistent with the Born rule and the projection postulate~\cite{AertsSassoli2014c}.

We will not describe here the details of this `hidden-measurement mechanism', as this is not the scope of this article. We only emphasize that its functioning requires the point particle representative of the state to move from the surface to the interior of the sphere, then back to the surface, thus implicitly ascribing the  
status of pure states also to the density operators. 

In other terms, if we take seriously the extended Bloch representation, we can say in retrospect that a key obstacle in our understanding of quantum measurements is that SQM was not considering all the possible pure states that can describe the condition of a physical entity, and that the missing ones were precisely those located inside of the generalized Bloch sphere, i.e., the density operators.

Our final and 
in a sense
most important argument in favour of the `density operators are pure states' interpretation is about showing that within the extended Bloch formalism a composite entity in an entangled state is naturally described as a system formed by two components that always remain in well defined states, precisely corresponding to the reduced states (\ref{individual states}), plus a third `element of reality' describing their non-local correlation. 
For this, we start by observing that (\ref{entanglement-operator}) can be written as~\cite{AertsSassoli2014c}:
\begin{equation}
D_\psi={1\over N}\left(\mathbb{I} +c_N\, {\bf r}\cdot\mbox{\boldmath$\Lambda$}\right),
\label{Bloch vector}
\end{equation}
where the real unit vector ${\bf r}$ is the representative of the ray-state $D_\psi$ within the generalized Bloch sphere $B_1(\real^{N^2-1})$, $c_N= ({N(N-1)\over2})^{1\over 2}$, and the components of the operator-vector \mbox{\boldmath$\Lambda$} are (a determination of) the generators of $SU(N)$, the \emph{special unitary group of degree $N$}, which are self-adjoint, traceless matrices obeying ${\rm Tr}\, \Lambda_i\Lambda_j = 2\delta_{ij}$, $i,j\in \{1,\dots, N^2-1\}$, forming a basis, together with the identity operator $\mathbb{I}$, for all the linear operators on ${\cal H}={\mathbb C}^N$.

In the same way, with ${\cal H}^A={\mathbb C}^{N_A}$, ${\cal H}^B={\mathbb C}^{N_B}$, $N=N_AN_B$, we can define the Bloch vectors: ${\bf r}^A,{\bf s}^A, \overline{\bf r}^A\in B_1(\real^{N_A^2-1})$, ${\bf r}^B,{\bf s}^B, \overline{\bf r}^B\in B_1(\real^{N_B^2-1})$, representative of the states $D_{\psi}^A$, $D_{\phi}^A$, $D^A$, $D_{\psi}^B$, $D_{\phi}^B$, $D^B$, respectively, by: $D_{\psi}^A={1\over N_A}(\mathbb{I}^A +c_{N_A}\, {\bf r}^A\cdot\mbox{\boldmath$\Lambda$}^A)$, $D_{\phi}^A={1\over N_A}(\mathbb{I}^A +c_{N_A}\, {\bf s}^A\cdot\mbox{\boldmath$\Lambda$}^A)$, $D^A={1\over N_A}(\mathbb{I}^A +c_{N_A}\, \overline{\bf r}^A\cdot\mbox{\boldmath$\Lambda$}^A)$, $D_{\psi}^B={1\over N_B}(\mathbb{I}^B +c_{N_B}\, {\bf r}^B\cdot\mbox{\boldmath$\Lambda$}^B)$, $D_{\phi}^B={1\over N_B}(\mathbb{I}^B +c_{N_B}\, {\bf s}^B\cdot\mbox{\boldmath$\Lambda$}^B)$, $D^B={1\over N_B}(\mathbb{I}^B +c_{N_B}\, \overline{\bf r}^B\cdot\mbox{\boldmath$\Lambda$}^B)$, where the $\Lambda_i^A$ are the $N^2_A-1$ generators of $SU(N_A)$, the $\Lambda_j^B$ are the $N^2_B-1$ generators of $SU(N_B)$, and $\mathbb{I}^A$ and $\mathbb{I}^B$ are the identity operators on ${\cal H}^A$ and ${\cal H}^B$, respectively. Note that for $N=2$, the components $\Lambda_i$ in (\ref{Bloch vector}) are simply the Pauli matrices, and ${\bf r}$ is a vector in the usual three-dimensional Bloch sphere. 

At this point, we observe that it is possible to use the remarkable property that the trace of a tensor product is the product of the traces, to construct a determination of the $SU(N)$ generators in terms of tensor products of the generators of $SU(N_A)$ and $SU(N_B)$. More precisely, defining the $N^2$ self-adjoint $N\times N$ matrices:
\begin{equation}
\Lambda_{(i,j)} = {1\over\sqrt{2}}\, \Lambda^{A}_{i}\otimes \Lambda^{B}_{j},\nonumber
\label{tensor-generators}
\end{equation}
where $ i=0,\dots,N_A^2-1$, $j=0,\dots,N_B^2-1$, and we have defined ${\Lambda^{A}_0}= ({2\over N_A})^{1\over 2}\mathbb{I}^A$ and ${\Lambda^{B}_0}= ({2\over N_B})^{1\over 2}\mathbb{I}^B$, it is easy to check that, apart $\Lambda_{(0,0)}=({2\over N})^{1\over 2}\mathbb{I}$, the remaining $N^2-1$ matrices are all traceless, mutually orthogonal and properly normalized, and therefore constitute a bona fide determination of the generators of $SU(N)$ that can be used in (\ref{Bloch vector}), to express the components of the vector ${\bf r}$, representative of the composite entity's state.

Using the orthogonality of the generators, the components of ${\bf r}$ are given by: $r_i=e_{N}{\rm Tr}\, D_\psi\Lambda_i, \quad i=1,\dots,N^2-1$, with $e_{N}= {N\over 2c_{N}}$, and similarly for the components of the Bloch vectors representing the sub-entities' states. With a direct calculation, one can then show that the entangled state ${\bf r}$ is of the tripartite direct sum form:
\begin{equation}
{\bf r} = d_{N_A}\overline{\bf r}^{A}\oplus d_{N_B}\overline{\bf r}^{B}\oplus {\bf r}^{\rm corr}.
\label{direct sum}
\end{equation}
where we have defined $d_{N_A}=({N_A-1\over N-1})^{1\over 2}$ and $d_{N_B}=({N_B-1\over N-1})^{1\over 2}$. In (\ref{direct sum}), the vector $\overline{\bf r}^{A}= a_1^2 {\bf r}^{A} + a_2^2 {\bf s}^{A}$ belongs to the one-entity Bloch sphere $B_1(\real^{N_A^2-1})$, and describes the state of $S^A$, whereas the vector $\overline{\bf r}^{B}= a_1^2 {\bf s}^{B} + a_2^2 {\bf r}^{B}$ belongs to the one-entity Bloch sphere $B_1(\real^{N_B^2-1})$, and describes the state of $S^B$. On the other hand, ${\bf r}^{\rm corr}$ is the component of the state which describes the correlation between the two sub-entities, and is of the form:
\begin{equation}
{\bf r}^{\rm corr} = d_{N_A,N_B}\overline{\bf r}^{AB} + {\bf r}^{\rm int},
\label{correlations}
\end{equation}
where $\overline{\bf r}^{AB}=a_1^2 {\bf r}^{AB}_1 + a_2^2{\bf r}^{AB}_2\in B_1(\real^{(N_A^2-1)(N_B^2-1)})$ is a vector with components 
$[\overline{\bf r}^{AB}]_{(i,j)}=a_1^2 [{\bf r}^{AB}_1]_{(i,j)} + a_2^2[{\bf r}^{AB}_2]_{(i,j)}$, with $[{\bf r}^{AB}_1]_{(i,j)}={\bf r}^{A}_i{\bf s}^{B}_j$, $[{\bf r}^{AB}_2]_{(i,j)}={\bf s}^{A}_i{\bf r}^{B}_j$, $i=1,\dots N_A^2-1$, $j=1,\dots N_B^2-1$, $d_{N_A,N_B}=({(N_A-1)(N_B-1)\over N-1})^{1\over 2}$, and the vector ${\bf r}^{\rm int}\in \real^{(N_A^2-1)(N_B^2-1)}$ describes the interference contribution (\ref{interference-contribution}). 

If the first two one-entity generators are chosen to be: $\Lambda^{A}_1 = |\psi^{A}\rangle\langle \phi^{A}| + |\phi^{A}\rangle\langle \psi^{A}|$, $\Lambda^{A}_2 =-i(|\psi^{A}\rangle\langle \phi^{A}| - |\phi^{A}\rangle\langle \psi^{A}|)$, $\Lambda^{B}_1 = |\psi^{B}\rangle\langle \phi^{B}| + |\phi^{B}\rangle\langle \psi^{B}|$, $\Lambda^{B}_2 =-i(|\psi^{B}\rangle\langle \phi^{B}| - |\phi^{B}\rangle\langle \psi^{B}|)$, ${\bf r}^{\rm int}$ only has four non-zero components, which for a suitably chosen order for the joint-entity generators are:
\begin{eqnarray}
{\bf r}^{\rm int} =e_N \sqrt{2}a_1a_2(\cos \alpha,\cos \alpha,-\sin \alpha,\sin \alpha, 0,\dots,0).
\label{interf vector}
\end{eqnarray}

According to (\ref{direct sum}), and different from the SQM formalism, we see that the extended Bloch representation allows to describe an entangled state as a ``less tangled'' condition in which the two sub-entities are always in the well-defined states $\overline{\bf r}^{A}$ and $\overline{\bf r}^{B}$, belonging to their respective one-entity Bloch spheres, which are clearly distinguished from their correlation, described by the vector (\ref{correlations}), which cannot be deduced from the states of the two sub-entities, in accordance with the general principle that the whole is greater than the sum of its parts (so that the states of the parts cannot generally determine the state of the whole).

We can observe that the interference contribution ${\bf r}^{\rm int}$ is what distinguishes the entangled state (\ref{entanglement})-(\ref{entanglement-operator}) from the \emph{separable state}: $D_\psi^{\rm sep}=a_1^2\, D_{\psi}^A \otimes D_{\phi}^B + a_2^2\, D_{\phi}^A\otimes D_{\psi}^B$. However, even when the interference contribution is zero, the separable (but non-product) state $D_\psi^{\rm sep}$ does not describe a situation of two experimentally separated entities, as is clear that the state vector $\overline{\bf r}^{A}$ is not independent from the state vector $\overline{\bf r}^{B}$, since their components both contain the parameters $a_1^2$ and $a_2^2$. Also, the components of $\overline{\bf r}^{AB}$ cannot be deduced from the knowledge of the components of $\overline{\bf r}^{A}$ and $\overline{\bf r}^{B}$, which means that a separable state is not a separated state, but a state that still describes a situation where the whole is greater than the sum of its parts.

Of course, when $a_2=0$ (or $a_1=0$), we are back to the situation of a so-called \emph{product state}. This manifests at the level of the Bloch representation in the fact that ${\bf r} = d_{N_A}{\bf r}^{A}\oplus d_{N_B}{\bf s}^{B}\oplus d_{N_A,N_B}{\bf r}_1^{AB}$, with the two sub-entity states ${\bf r}^{A}$ and ${\bf s}^{B}$ now totally independent from one another, and able to fully determine the joint-entity contribution ${\bf r}_1^{AB}$, so that there are no genuine emergent properties in this case.

Therefore, considering the general description (\ref{direct sum}), and the previously mentioned arguments in favor of the `density operators are pure states' interpretation, we are now in a position to formulate a completed quantum mechanical principle, in replacement of the SQMP1, which can restore the full intelligibility of entangled states:

\vspace{0.10cm}
\noindent {\bf CQMP1} \emph{If $S$ is a physical entity with Hilbert space ${\cal H}$, then each density operator of ${\cal H}$ is a pure state of $S$ and all the pure states of $S$ are of this kind.}
\vspace{0.10cm}

Of course, the CQMP1 does not imply that a density operator cannot be used to also describe a situation of subjective ignorance of the experimenter, regarding the pure state of an entity. It simply means that within the quantum formalism a same mathematical object can be used to model different situations, which are not easy to experimentally distinguish, because of the linearity of the trace used to calculate the transition probabilities. 

However, a distinction between `pure state-density operators' and `mixed state-density operators' is in principle possible, for instance if one can set up an experimental context producing a non-linear evolution of the states, as in this case mixtures and pure states will evolve in a different way, and their ontological difference can become observable (see~\cite{AertsSassoli2014c} for some additional considerations regarding the distinguishability of pure and mixed states in a measurement context).

It is worth mentioning that a completed quantum mechanics retaining all the principles of SQM, apart the SQM1 which is to be replaced by the CQMP1, was already proposed by one of us many years ago~\cite{Aerts2000}. At the time the proposal was motivated by the existence of a mechanistic classical laboratory situations able to violate Bell's inequalities exactly as quantum entities in EPR-experiments can do~\cite{Aerts1991}. Today, considering that the `density operators are pure states' interpretation is an integral part of the extended Bloch representation, which provides a possible solution to the measurement problem~\cite{AertsSassoli2014c} and, as we have shown in this article, also allows to obtain a partitioning of the entangled states where their correlative aspects remain clearly and naturally ``disentangled'' from the description of the sub-entities' states, we believe that the proposal has reached the status of a firmly founded scientific hypothesis, only waiting for an experimental confirmation.

A last remark is in order. Even in the simplest case of two entangled qubits ($N_A=N_B=2$), where the two one-entity states $\overline{\bf r}^{A}$ and $\overline{\bf r}^{B}$ can be represented within our 3-dimensional Euclidean space (for instance as directions, in the case of spins), the correlation vector ${\bf r}^{\rm corr}$ is already 9-dimensional, and therefore is no longer describable within our Euclidean theatre. This is in accordance with the observed non-local effects that are produced by entangled entities, which are insensitive to spatial separation, and which therefore should be understood as effects resulting from the existence of genuinely \emph{non-spatial} correlations. 

In other terms, the solution we have proposed to the second entanglement paradox, via the extended Bloch representation, also suggests that non-locality should be understood as a manifestation of the 
non-spatial
nature of quantum entities. This means that our approach also offers a possible solution to the first entanglement paradox, as is clear that if quantum entities are non-spatial entities, then their interconnections, when in entangled states, need not to happen through space, and therefore can remain perfectly insensitive to spatial separation.


\begin{thebibliography}{}

\bibitem{Schroedinger1935} E. Schr{\oe}dinger, Naturwissenschaftern 23, 807 (1935). English translation: John D. Trimmer, Proceedings of the American Philosophical Society, 124, 323 (1980). Reprinted in: J. A. Wheeler and W. H. Zurek (Eds.), \emph{Quantum Theory and Measurement} (Princeton University Press, Princeton, 1983) 152.
\bibitem{Fraassen1991} B. C. Van Fraassen, \emph{Quantum Mechanics: An Empiricist View}, Oxford University Press, Oxford, New York, Toronto (1991).
\bibitem{Beltrametti1981} E. Beltrametti and G. Cassinelli, \emph{The Logic of Quantum Mechanics}, Reading, Mass.: Addison-Wesley (1981).
\bibitem{Aerts2000} D. Aerts, ``The description of joint quantum entities and the formulation of a paradox," Int. J. Theor. Phys. 39, 485--496 (2000).
\bibitem{Hughston1993} L. P. Hughston, R. Jozsa and William K. Wootters,``A complete classification of quantum ensembles having a given density matrix,'' Physics Letters A 183, 14--18 (1993).
\bibitem{AertsSassoli2014c} D. Aerts and M. Sassoli de Bianchi, ``The Extended Bloch Representation of Quantum Mechanics and the Hidden-Measurement Solution to the Measurement Problem,'' Annals of Physics 351, 975--1025 (2014).
\bibitem{Bloch1946} F. Bloch, ``Nuclear induction,'' Phys. Rev. 70, 460--474 (1946).
\bibitem{Kimura2003} G.  Kimura,``The Bloch Vector for $N$-Level Systems,'' Phys. Lett. A 314, 339 (2003).
\bibitem{Aerts1991} D. Aerts, ``A mechanistic classical laboratory situation violating the Bell inequalities with $2\sqrt{2}$, exactly `in the same way' as its violations by the EPR experiments,'' Helv. Phys. Acta 64, 1--23 (1991).



\end{thebibliography}
\end{document}